\documentclass[a4paper,fleqn,usenatbib]{mnras}

\usepackage{newtxtext,newtxmath}
\usepackage{xcolor}

\usepackage[T1]{fontenc}
\usepackage{ae,aecompl}

\usepackage{graphicx}	
\usepackage{amsmath}	
\usepackage{amssymb}	

\title[$H_0$ tension relieved]{Gaia Cepheid parallaxes and `Local Hole' relieve $H_0$ tension}

\author[T. Shanks et al.]{
T. Shanks,\thanks{E-mail: tom.shanks@durham.ac.uk (TS)}
L.M. Hogarth,
N. Metcalfe.\\
Physics Department, Durham University, South Road, Durham, DH1 3LE, England
}

\date{Accepted 2018 December 17. Received 2018 December 17; in original form 2018 October 5}

\pubyear{2015}

\begin{document}
\label{firstpage}
\pagerange{\pageref{firstpage}--\pageref{lastpage}}
\maketitle

\begin{abstract}

There is an $\approx9\pm2.5$\% tension between the value of Hubble's
Constant, $H_0=67.4\pm0.5$km\,s$^{-1}$Mpc$^{-1}$, implied by the {\it
Planck} microwave background power spectrum and that given by the
distance scale of $H_0=73.4\pm1.7$km\,s$^{-1}$Mpc$^{-1}$. But with  a
plausible assumption about a {\it Gaia} DR2 parallax systematic offset,
we find that {\it Gaia} parallax distances of Milky Way Cepheid
calibrators are $\approx12-15$\% longer than previously estimated.
Similarly, {\it Gaia} also implies $\approx4.7\pm1.7$\% longer distances
for 46 Cepheids than previous distances on the scale of Riess et al.
Then we show that the existence of an $\approx150$h$^{-1}$Mpc `Local
Hole' in the galaxy distribution implies an outflow of
$\approx500$km\,s$^{-1}$. Accounting for this in the recession
velocities of SNIa standard candles out to $z\approx0.15$ reduces $H_0$
by a further $\approx1.8$\%. Combining the above  two results would
reduce the distance scale $H_0$ estimate by $\approx7$\% from
$H_0\approx73.4\pm1.7$ to $\approx68.9\pm1.6$ km\,s$^{-1}$Mpc$^{-1}$, in
reasonable agreement with the {\it Planck} value. We conclude that the
discrepancy between distance scale and {\it Planck} $H_0$ measurements remains
unconfirmed due to uncertainties caused by {\it Gaia} systematics and an
unexpectedly inhomogeneous local galaxy distribution.

															

\end{abstract}

\begin{keywords}
cosmology -- distance scale -- Hubble's Constant
\end{keywords}



\section{Introduction}

The history of measuring the Hubble Constant via the distance scale has
been one of contention, with Hubble's original value of
$H_0\approx500$km\,s$^{-1}$Mpc$^{-1}$ gradually reducing to the present
$H_0\approx73$km\,s$^{-1}$Mpc$^{-1}$. The problem has been that to estimate
$H_0$ accurately we need to go to distances beyond the largest local
inhomogeneities so that the recession velocity completely dominates any
peculiar velocity and unfortunately this is beyond the reach of
geometric distance indicators such as parallax	and even primary
distance indicators such as Cepheids.	

Recently, there has been a tension noted between the $H_0$ estimated from
models fitted to the acoustic peaks in the Planck CMB power spectrum
which give $H_0=67.4\pm0.5$km\,s$^{-1}$Mpc$^{-1}$ \citep{Planck2018} and
the distance scale estimates which give
$H_0=73.4\pm1.7$km\,s$^{-1}$Mpc$^{-1}$. This $\approx9$\% discrepancy is now at
the $3.5\sigma$ level and is regarded as a serious tension in the Hubble
parameter \citep{Riess2016}.

Although there remains the possibility that the uncertainties have been
underestimated in {\it both} the distance scale and {\it Planck} $H_0$ results
(see eg \citealt{Feeney2018}), here we focus on  two developments that
may act to reduce distance scale estimates of $H_0$. The first comprises
new parallax distances to Milky Way Cepheids from {\it Gaia} DR2
\citep{GaiaDR2_2018}. We compare these to previous Cepheid
parallax distances and also to main-sequence fitted distances for
Cepheids in Galactic open clusters.  We similarly compare directly 
to Cepheids on the distance scale of \cite{Riess2018a}, calibrated 
by parallax and two other geometric methods.

Second, we review the evidence for the `Local Hole' from the work of 
\cite{Whitbourn2014,Whitbourn2016} and references therein. We then estimate the effect
of the resulting outflow on scales of $\approx150$h$^{-1}$Mpc on the redshifts 
of SNIa that are used in the Hubble diagram by fitting for $H_0$ and $\Omega_m$.


\begin{figure}
	\includegraphics[width=\columnwidth]{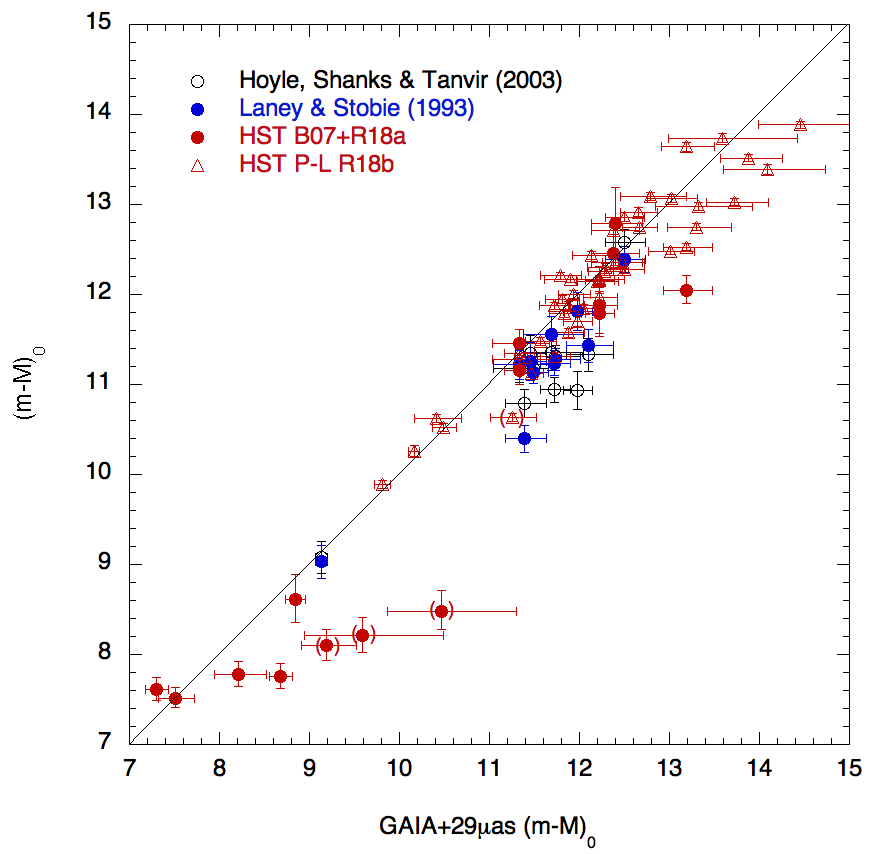}
    \caption{Comparison between Cepheid distances based on {\it Gaia} parallaxes (assuming the 
    29 $\mu$as `quasar' correction), compared to the HST parallaxes used by \protect\cite{Riess2018a}
    to  help zeropoint their Cepheid scale. Also shown is the same {\it Gaia} comparison for 
    Cepheids in open clusters  with main sequence fitted distances 
    from \protect\cite{Laney1993} and  \protect\cite{Hoyle2003}. Finally, 
    the distance moduli of the 46 Cepheids of \protect\cite{Riess2018b} 
    are also compared to those from {\it Gaia}. The Cepheids $l$ Car, W Sgr, RT Aur and T Mon are 
    also plotted but with brackets  because they were left out of numerical comparisons (see text).  
    }
    \label{fig:gaia}
\end{figure}

\begin{table*}
\begin{center}
\begin{tabular}{ccccccc}
\hline
Cepheid   & Ref   &HST $\pi$          &  HST $(m-M)_0$  & {\it Gaia} $\pi$     & {\it Gaia} $(m-M)_0$& {\it Gaia} G\\
          &       &  (mas)            &                 &(mas)           &($29\mu$as corr.)& (mag)\\ 
\hline
SS CMA    & R18a  & 0.389$\pm$0.029   & 12.05$\pm$0.16 &0.201$\pm$0.029 & 13.19$\pm$0.28 &  9.52\\
XY Car    & R18a  & 0.438$\pm$0.047   & 11.79$\pm$0.23 &0.330$\pm$0.027 & 12.23$\pm$0.16 &  8.94\\ 
VX Per    & R18a  & 0.420$\pm$0.074   & 11.88$\pm$0.39 &0.330$\pm$0.031 & 12.23$\pm$0.19 &  8.86\\
VY Car    & R18a  & 0.586$\pm$0.044   & 11.16$\pm$0.16 &0.512$\pm$0.041 & 11.33$\pm$0.17 &  7.33\\
WZ Sgr    & R18a  & 0.512$\pm$0.037   & 11.45$\pm$0.16 &0.513$\pm$0.077 & 11.33$\pm$0.31 &  7.65\\
S Vul     & R18a  & 0.322$\pm$0.040   & 12.46$\pm$0.27 &0.305$\pm$0.041 & 12.38$\pm$0.27 &  8.05\\
X Pup     & R18a  & 0.277$\pm$0.047   & 12.79$\pm$0.37 &0.302$\pm$0.043 & 12.40$\pm$0.28 &  8.30\\
$l$ Car$^*$& B07  & 2.010$\pm$0.20    &  8.48$\pm$0.22 &0.777$\pm$0.256 & 10.47$\pm$0.72 &  3.79(DR1)\\
$\zeta$ Gem& B07  & 2.780$\pm$0.18    &  7.78$\pm$0.14 &2.250$\pm$0.301 &  8.21$\pm$0.29 &  4.06\\
$\beta$ Dor& B07  & 3.140$\pm$0.16    &  7.52$\pm$0.11 &3.112$\pm$0.284 &  7.52$\pm$0.20 &  4.10(DR1)\\
W Sgr$^*$ & B07   & 2.280$\pm$0.20    &  8.21$\pm$0.19 &1.180$\pm$0.412 &  9.59$\pm$0.77 &  5.1(B07)\\
X Sgr     & B07   & 3.000$\pm$0.18    &  7.61$\pm$0.13 &3.431$\pm$0.202 &  7.30$\pm$0.13 &  4.32\\
FF Aql    & B07   & 2.810$\pm$0.18    &  7.76$\pm$0.14 &1.810$\pm$0.107 &  8.68$\pm$0.13 &  5.14\\
T Vul     & B07   & 1.900$\pm$0.23    &  8.61$\pm$0.26 &1.674$\pm$0.089 &  8.84$\pm$0.11 &  5.44\\
RT Aur$^*$& B07   & 2.400$\pm$0.19    &  8.10$\pm$0.17 &1.419$\pm$0.203 &  9.20$\pm$0.31 &  5.9(B07)\\
\hline
\end{tabular}
\end{center}
\caption{Cepheid distance moduli estimated from {\it Gaia} parallaxes compared
to previous HST parallax data  data. B07 represents the Hubble Space
Telescope (HST) FGS Cepheid parallaxes of \protect\cite{Benedict2007}.
R18a represents the HST WFC3 Cepheid parallaxes of
\protect\cite{Riess2018a}. {\it Gaia} parallaxes, $\pi$, are listed
uncorrected for systematic offset whereas {\it Gaia} distance moduli are based
on parallaxes corrected by adding $29\mu$as. G magnitudes are from {\it Gaia} DR2
except for $l$ Car and $\beta$ Dor where we prefer DR1 G magnitudes and W
Sgr and RT Aur where we prefer the V magnitude of B07 converted to G
using $G\approx V+0.4$. $^*$ denotes FGS stars rejected on grounds of {\it Gaia}
saturation and high parallax fractional error. The remainder all have {\it Gaia} parallax 
fractional error in the range $5-17$\%.
}
\label{table:gaia_hst}
\end{table*}

\section{New Cepheid parallax distances from Gaia.}
\label{sec:cepheid}
{\it Gaia} DR2 has provided parallaxes of unprecedented statistical accuracy
to many hundreds of Cepheid variables. \cite{Riess2018b} have analysed 46 of
these where HST photometry exists and have concluded that their previous
Cepheid P-L relation zeropoints and distance scale have been confirmed,
leaving the Hubble Constant at $H_0=73.4\pm1.7$km\,s$^{-1}$Mpc$^{-1}$.
However, as \cite{Riess2018b} note, there are systematic uncertainties
in the {\it Gaia} DR2 data particularly in the parallax zeropoint and the
effects of saturation that affect Cepheid distances.
\cite{Lindegren2018} show that WISE quasars have an average {\it Gaia}
parallax of -29 $\mu$as when it should be zero. They warn that this
offset may not be constant with sky position, star colour or magnitude.
They suggest that the offset could be fitted out in individual star
samples. This is the approach used by \cite{Riess2018b} who found an
average offset of -46 $\mu$as against previous P-L based distances
to 46 Cepheids (see below). Similarly, \cite{Zinn2018} report an offset of
$-52.8\pm2.4\pm1$(syst.)$\mu$as on astroseismological/APOGEE
spectroscopic distances to stars in the Kepler field.  A range of
offsets is also quoted in Table 1 of \cite{Arenou2018}.

Here, we shall assume as our baseline, the average -29$\mu$as offset
from the quasars,  estimated by  \cite{Lindegren2018} for the {\it Gaia}
collaboration, and add 29$\mu$as to correct our {\it Gaia} parallaxes.
Clearly the difference between this average value and, for example, the
value of \cite{Zinn2018} emphasises  the possibility that the offset may
vary with sky position or another parameter. But the reason we prefer
the 'quasar' offset is that its ideal `model'  parallax of $\pi=0$ is
indisputable unlike almost all other distance comparisons quoted by
\cite{Arenou2018} and \cite{Zinn2018} (see also \citealt{Stassun2018}).
Unfortunately, the colour and magnitude ranges of quasars
are too small to base reliable corresponding corrections for Cepheid
parallaxes. But we have checked that adjusting the {\it Gaia}
parallaxes for the possible dependence on ecliptic latitude given in
Fig. 7 (right) of \cite{Lindegren2018} changes our results
insignificantly.


The non-linear $r=1/\pi$ relation between distance, $r$, and parallax,
$\pi$, can also cause statistical and systematic bias in parallax
distances. \cite{Luri2018} has made a thorough review of these effects
for {\it Gaia} parallaxes, including approaches such as those of
\cite{Smitheichhorn1996, LK1973, BJ2015}. However, our fractional errors
in {\it Gaia} parallax are generally $\approx10$\% (see Tables
\ref{table:gaia_hst}, \ref{table:gaia_open}) and these authors agree
that this makes them less susceptible to statistical bias. Indeed, tests
on the samples used here assuming the average error/parallax ratio  and cut at $<20$\%
all gave $<0.5\mu$as bias.  Therefore we adopt a simple approach and
compare distances from {\it Gaia} parallaxes directly with previous
distance measurements.

We  first consider 3 samples on which the P-L relation has traditionally been
calibrated. Two of these comprise the parallax Cepheid samples discussed by
\cite{Riess2018b}, namely the 10 parallax stars from HST FGS
measurements of \cite{Benedict2007} and the 7 HST WFC3 parallax stars of
\cite{Riess2018a} (see Table \ref{table:gaia_hst}). Similar to
\cite{Riess2018b} we have excluded $\delta$ Cep and Y Sgr because they show
negative  parallaxes in {\it Gaia} DR2, possibly due to data corruption. To
these we add the 14 Cepheids in 11 open clusters whose distances were
obtained by main sequence fitting by \cite{Laney1993} and
\cite{Hoyle2003} who added NIR K band photometry to try and improve their
reddening estimates and hence the distances (see Table \ref{table:gaia_open}).

\begin{table*}
\begin{center}
\begin{tabular}{ccccccc}
\hline
Open     & Cepheid(s)        &     $(m-M)_0$    &   $(m-M)_0$          & {\it Gaia} $\pi$     &{\it Gaia} $(m-M)_0$  & {\it Gaia} G\\
Cluster  &                   &(Laney \& Stobie 1993)&  (Hoyle et al. 2003)       &(mas)           &($29\mu$as corr.)& (mag)\\ 
\hline
NGC6649  & V367 Sct          & 11.28  & 11.31$\pm$0.12      &0.4203$\pm$0.053 & 11.74$\pm$0.26 &10.50\\
M25      & U Sgr             &  9.03  &  9.08$\pm$0.18      &1.4601$\pm$0.045 &  9.14$\pm$0.07 & 6.50\\ 
NGC6664  & EV Sct            & 10.40  & 10.79$\pm$0.15      &0.4969$\pm$0.054 & 11.40$\pm$0.22 & 9.64\\
WZ Sgr   & WZ Sgr            & 11.22  & 11.18$\pm$0.16      &0.5131$\pm$0.077 & 11.33$\pm$0.31 & 7.65\\
Lynga 6  & TW Nor            & 11.43  & 11.33$\pm$0.18      &0.3505$\pm$0.045 & 12.10$\pm$0.26 &10.50\\
NGC6067  & QZ Nor, V340      & 11.13  & 11.18$\pm$0.12      &0.4744$\pm$0.038 & 11.49$\pm$0.16 & 8.62\\
VdB 1    & CV Mon            & 11.26  & 11.34$\pm$0.21      &0.4823$\pm$0.041 & 11.46$\pm$0.17 & 9.60\\
Tr 35    & RU Sct            & 11.56  & 11.36$\pm$0.20      &0.4307$\pm$0.070 & 11.70$\pm$0.34 & 8.81\\
NGC6823  & SV Vul            & 11.81  & 10.93$\pm$0.21      &0.3729$\pm$0.030 & 11.98$\pm$0.16 & 6.87\\
NGC129   & DL Cas            & 11.24  & 10.94$\pm$0.14      &0.4222$\pm$0.034 & 11.73$\pm$0.16 & 8.58\\
NGC7790  & CF, CEa, CEb Cas  & 12.39  & 12.58$\pm$0.14      &0.2871$\pm$0.032 & 12.50$\pm$0.22 &10.73\\
\hline
\end{tabular}
\end{center}
\caption{Cepheid distance moduli estimated from {\it Gaia} parallaxes
compared to previous distance moduli estimated from main-sequence
fitting for Cepheids in open clusters. We shall assume the Hoyle et al
uncertainties also apply to the Laney \& Stobie distance moduli.  {\it
Gaia} parallaxes, $\pi$, are listed uncorrected for systematic offset
whereas {\it Gaia} distance moduli are based on parallaxes corrected by
$+29\mu$as.  These Cepheids all have {\it Gaia} parallax
fractional errors in the range 3-17\%. Note that the association of SV
Vul with NGC6823 is controversial (see \protect\citealt{Anderson2013}).
}
\label{table:gaia_open}
\end{table*}

The comparison of these previous distance moduli with the {\it Gaia} DR2
parallax distances with the $29\mu$as `quasar' correction added
to form the corrected distance moduli, $(m-M)_0$, are
shown in Fig. \ref{fig:gaia}. We see that there is a significant
discrepancy that seems independent of distance. Conservatively, we
compare with the \cite{Laney1993} distances to the open clusters - these
are larger overall than the \cite{Hoyle2003} alternatives and may be
more considered the previous standard values. Note that the
$29\mu$as correction only falls to $<1$\% of the parallax for 
$(m-M)_0<7.7$mag ie only the {\it Gaia} parallaxes for FGS Cepheids of
\cite{Benedict2007} are relatively immune to this systematic.

With the quasar offset and further excluding $l$ Car, W Sgr and RT Aur
of \cite{Benedict2007} on the grounds they show the lowest
parallax/error ratios as well as having saturated {\it Gaia} magnitudes
with $G<6$mag, the 23 remaining {\it Gaia} moduli are 
$0.30\pm0.06$mag longer than the previous distance moduli or
$14.8\pm3$\% greater in distance. Here and in
what follows we have minimised $\chi^2$ on the distance moduli
differences with respect to the {\it Gaia} and the other sample's errors
(see Tables \ref{table:gaia_hst},\ref{table:gaia_open}) combined in
quadrature. Further excluding the 5 remaining FGS stars with $G<6$ in
case their parallaxes are affected by saturation, the 18 {\it Gaia}
distance moduli left are  $0.32\pm0.06$mag longer than the
previous moduli or $15.9\pm3$\% greater in distance.
Alternatively, excluding the 11 Cepheid open clusters of
\cite{Laney1993} the remaining 7 WFC3 and 5 FGS Cepheids with HST
parallaxes (still excluding l Car, W Sgr and RT Aur) the {\it Gaia}
distance moduli are $0.25\pm0.08$mag higher than previously
or $12.2\pm3.8$\% longer in distance. For just the 7 WFC3 stars the
{\it Gaia} moduli are $0.28\pm0.12$mag higher or
$13.8\pm5.7$\% longer in distance. Note that for their distance scale,
\cite{Riess2018a} report that  their calibration route via Milky Way
Cepheid parallaxes produces $4.8\pm3.3$\% longer distances than the
NGC4258 megamaser and eclipsing binaries routes combined. This partly
explains why we further find that the final distance moduli reported for
these 7 WFC3 Cepheids in Table 2 of \cite{Riess2018a} are
$0.16\pm0.08$mag  or $7.6\pm3.8$\% longer than their HST parallax
distances. Lastly, we note that each of the above differences remain
significant if we assume the $46\pm6\mu$as parallax correction of 
\cite{Riess2018b} rather than  $29\mu$as e.g for
all 23 previous calibrators the {\it Gaia} moduli are now
$0.24\pm0.07$mag longer c.f. $0.30\pm0.06$mag.


The results of Fig. \ref{fig:gaia} as discussed so far are based on a
comparison of {\it Gaia} parallaxes with  HST parallax and main-sequence
fitted distances to the Milky Way Cepheids previously used to calibrate
the  P-L relation. These give important contextual evidence that the
{\it Gaia} distances may imply a longer Cepheid scale. But the most
direct comparison with the scale of \cite{Riess2018b} comes from their 
sample of 46 Cepheids, also shown in Fig. \ref{fig:gaia}. We follow
these authors by excluding the Cepheid T Mon, bracketted in Fig.
\ref{fig:gaia}. We also note that 4/46 Cepheids have parallax fractional
errors $>20$\%. \cite{Riess2018b} found the {\it Gaia}
parallax results supported the previous distance scale of
\cite{Riess2018a}. However,  \cite{Riess2018b} left the
{\it Gaia} parallax offset as a free parameter in comparing to the
distances of 46 Cepheids with photometric distances from their P-L
relation. Assuming no distance dependence (i.e. their $\alpha=1$), they
found a parallax offset of  $-46\pm6\mu$as  (see their Fig. 5) compared
to the $-29\mu$as adopted here. If we assume the $29\mu$as `quasar'
correction and  a median {\it Gaia} parallax of $0.33\pm0.029$mas for
their 46 Cepheids then this would imply  a $4.7\pm1.7$\% increase in
their distance scale. If we assumed a zero {\it Gaia} parallax offset
correction then this route would imply  a $13.9\pm1.8$\% increase.

We note that a direct comparison of {\it Gaia} ($+29\mu$as) and these 46
Cepheid `photometric' distance moduli (see Fig. \ref{fig:gaia})  gives a
smaller distance difference of $0.035\pm0.03$mag or $1.6\pm1.4$\%.
Correcting by $+46\mu$as in turn gives an $(m-M)_0$ difference in the
opposite direction of $-0.045\pm0.03$mag. Thus the total difference is
$0.08\pm0.03$mag or $3.75\pm1.4$\%, similar to  the $4.7\pm1.7$\%
derived from parallaxes directly.  Assuming zero correction, the {\it
Gaia} $(m-M)_0$ are longer by $0.175\pm0.03$mag or by $8.4\pm1.4$\% in
distance. Thus the basic result remains that varying the assumed {\it
Gaia} correction from zero to $0.029$mas gives larger {\it Gaia}
distances by $\approx2-8$\%. So, at one end of this range, this difference might
explain the discrepancy between the Cepheid parallax and
megamaser/eclipsing binary Cepheid zeropoints. At the other, the Cepheid
scale calibrated by {\it Gaia} parallaxes would become more compatible with
Planck. But until the systematic {\it Gaia} parallax offset becomes
better determined it may be  impossible to claim that {\it Gaia} has either
contradicted or confirmed the current Cepheid scale.




Summarising, assuming a $29\mu$as correction offset to {\it Gaia}
parallaxes, we find 12-15\% longer distances to 23 Cepheids previously
used as Cepheid P-L relation calibrators.  Only about half of this
increase might be expected based on difference between the
previous parallax and the other geometric calibrations of
\cite{Riess2018a}. {\it Gaia} ($+29\mu$as offset)
parallaxes for the 46 Cepheids of \cite{Riess2018b} indicate a
$4.7\pm1.7$\% increase in their Cepheid distances. This would imply
their $H_0\approx70.2\pm1.2$km\,s$^{-1}$Mpc$^{-1}$. However, this latter
result depends on the assumed {\it Gaia} parallax offset.
Assuming zero offset would give a  bigger $13.9\pm1.8$\% reduction in
the $H_0$ of \cite{Riess2018a}.

\section{$H_0$ and the `Local Outflow'}
\label{sec:outflow}
Next, we consider the peculiar velocity outflow caused by the `Local Hole'.
Evidence for a local underdensity on the scale of $\approx150h^{-1}$Mpc
goes back to the galaxy count data of \cite{Shanks1984} and has been
confirmed particularly in the Southern Galactic Cap in the 2dF Galaxy
Redshift Survey by \cite{Busswell2004} and in 2MASS counts  by
\cite{Frith2003, Frith2006}. More recently all sky redshift surveys
including SDSS and 6dFGS have provided further confirmation as discussed
by \cite{Keenan2012,Keenan2013, Whitbourn2014, Whitbourn2016}. Here we
follow \cite{Whitbourn2014} to estimate the effect of the `Local
Outflow'.

\begin{figure}
	\includegraphics[width=\columnwidth]{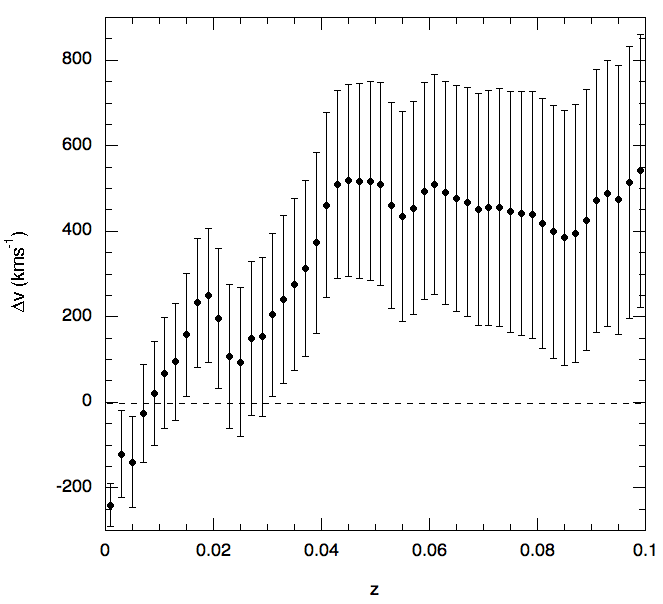}
	\caption{Outflow peculiar velocity, $\Delta v$ (km\,s$^{-1}$) inferred
	via  eqs. \ref{eq:density} and \ref{eq:linear} from the 
	6dFGS+SDSS galaxy redshift distributions in the 3 sky  areas of
	\protect\cite{Whitbourn2014}. Here, outflows have positive $\Delta v$. 
	}
    \label{fig:deltavz}
\end{figure}

We therefore use the three sky areas of Whitbourn et al within each of which the
$\delta\rho_g(r)/\bar{\rho}_g$ was estimated as a function of redshift by
dividing the observed $n(z)$ limited at $K<12.5$ to a homogeneous model.
We then form: 

\begin{equation}
\frac{\delta\rho_g(<r)}{\bar{\rho}_g}= \frac{1}{V(r)}\sum_{i} \bigg(\frac{dn}{n}\bigg)_i 4\pi r_i^2\delta r
\label{eq:density}
\end{equation}

\noindent where $\big(\frac{dn}{n}\big)_i$ are taken from averaging the data
shown in Fig. 3 (a, b, c) of \cite{Whitbourn2014}. $r_i$ are the corresponding comoving
distances, $\delta r$ is the comoving bin size and $V(r)$ is the
spherical volume to radius, $r$. We then apply linear theory to
relate the fractional velocity change to the galaxy overdensity,
$\delta\rho_g(r)/\bar{\rho}_g$:

\begin{equation}
\frac{\Delta v}{v}=-\frac{1}{3}\frac{\delta\rho_g(<r)}{\bar{\rho}_g} \frac{\Omega_m^{0.6}}{b}
\label{eq:linear}
\end{equation}

\noindent where $b$ is the galaxy bias. Here we assume $b=1$ as appropriate for $K$ 
selected galaxies in the standard cosmological model (e.g.\citealt{Whitbourn2014}).

Fig. \ref{fig:deltavz} shows the predicted outflow velocity from an area
weighted average over the 3 regions of \cite{Whitbourn2014}. This
averaging implies a spherically symmetric underdensity but clearly this
is only a rough approximation. In future work we shall explore the
effect of relaxing this assumption. At $z\approx0.05$ the ratio $\Delta
v/v$ peaks at $3.5$\% while  averaging $\approx1.8$\% in the range
$0.01<z<0.15$ used by \cite{Riess2016}. Here, we have
assumed that the only contribution to $\Delta v/v$ is from $z<0.1$,
leaving $\Delta v/v$ to decline towards zero in the $0.1<z<0.15$ range.
So for a local determination of $H_0$ measured in this range, the effect
of the `Local Hole' will tend to lower $H_0$ by 
$\approx1.8$\%.  This effect will reduce the
$H_0=73.4\pm1.7$km\,s$^{-1}$Mpc$^{-1}$ distance scale measurement of
\cite{Riess2018a} to $H_0\approx72.1\pm1.6$km\,s$^{-1}$Mpc$^{-1}$.

We next check the effect of this Local Hole outflow on the Hubble
Diagram to look for any inconsistency with the standard model fit. We
therefore apply the correction $\Delta v$ from Fig. \ref{fig:deltavz} to
obtain the corrected SNIa redshift, $z_{cor}$:

\begin{equation}
1+z_{cor}=\frac{1+z_{SNIa}}{1+\Delta v/c}.
\label{eq:zcor}
\end{equation}

\noindent These corrected redshifts are then assumed to calculate  the distance moduli of
the {\it Pantheon} sample of 1048 SN1a of \cite{Scolnic2018} according to,

\begin{equation}
m-M=25+5log_{10}((1+z_{cor})\times r).
\label{eq:dm}
\end{equation}



\noindent Here, we apply the average correction shown in
Fig. \ref{fig:deltavz}, irrespective of which sky area the SNIa	 is
located and data points are assumed uncorrelated. The
Hubble diagram for  1048  $z>0.01$ SNIa is then $\chi^2$ fitted for
$\Omega_m$ and $H_0$.  For the original sample with no corrections, the
best fit was $H_0=73.4\pm0.2$km\,s$^{-1}$Mpc$^{-1}$ and
$\Omega_m=0.28\pm0.01$ with $\chi^2=0.9902$. These are statistical
errors ignoring systematics. The best fit to the outflow corrected SNIa
data is $H_0=72.4\pm0.2$km\,s$^{-1}$Mpc$^{-1}$, close to the
$H_0=72.2\pm1.6$km\,s$^{-1}$Mpc$^{-1}$ estimated above, and
$\Omega_m=0.33\pm0.015$ with $\chi^2=0.9886$. Thus, including outflows
allows slightly lower values of $H_0$ and slightly higher
values of $\Omega_m$.







\section{Conclusions}
\label{sec:conclusions}
The most significant potential change to $H_0$ comes from the new
Cepheid parallaxes measured by {\it Gaia}. In comparisons with previous
Cepheid calibrators, we found  an average distance increase of
$\approx12-15$\%. However, there is still uncertainty here in that we
have assumed the 29$\mu$as  correction for the parallaxes. Although this
reduces the distances to Cepheids from the raw {\it Gaia} results,
\cite{Lindegren2018} have emphasised that the offsets may be sky
position and colour dependent. Indeed, the difference between our
conclusions and those of \cite{Riess2018b}, who found consistency with
the previous distance scale, is that \cite{Riess2018b} left this
systematic {\it Gaia} parallax offset a free parameter and fitted for it
in their sample of 46 Galactic Cepheids with {\it Gaia} parallaxes.
These authors found an offset of $-46\mu$as that, when corrected, gave a
best-fit distance scale of $1.006\pm0.033$ relative to their previous
scale. However, given that the offset is fitted, this is clearly not an
independent confirmation of the Cepheid scale. If we adopt instead our
$29\mu$as correction for their 46 Cepheids then this would imply a 
$4.7\pm1.7$\% increase in their distance scale. This corresponds to a
decrease in Hubble's Constant from
$H_0=73.4\pm1.7$km\,s$^{-1}$Mpc$^{-1}$ to
$H_0=70.2\pm1.2$km\,s$^{-1}$Mpc$^{-1}$. The bigger $\approx12$\%
increase in the distances to Cepheids with HST parallaxes could  then
more than reconcile their known $4.8\pm3.3$\% difference with the other
geometric Cepheid calibrators based on  eclipsing binaries and the
NGC4258 maser \citep{Riess2018a}. But much clearly depends on the value
of the {\it Gaia} parallax systematic offset. We acknowledge arguments
supporting the offset used by \cite{Riess2018b} from e.g.
\cite{Zinn2018}. Whether this current {\it Gaia} scale can compete with the
alternative Cepheid geometric calibrations awaits an improved {\it Gaia}
astrometric solution.

In terms of possible problems with the previous Cepheid calibrators, we
note that {\it Gaia} parallaxes have the advantage over HST parallaxes that
they are global, with no need of modelling background star distances. In
the case of main sequence distances, these fits assume a universal
Galactic reddening law but a spatial dependence  is increasingly
discussed \citep{Fitz2007,Anderson2013}.

Then considering the effect of  outflow due to the `Local Hole' we have
found that an $\approx1.8$\% decrease in average galaxy velocities out
to $z\approx0.15$ is likely when the effect of local underdensities are
taken into account. Here we have assumed linear theory in terms of
relating underdensity to outflow velocity which is an approximation but
others using more sophisticated models have come to similar conclusions
(e.g. \citealt{Hoscheit2018}). We have checked whether our
linear outflow model leads to any inconsistency in the SNIa Hubble
diagram using the data  of \cite{Scolnic2018} but as long as a slight
rise in $\Omega_m$ is allowed from $\Omega_m=0.28$ to 0.33, c.f.
$\Omega_m=0.315\pm0.0007$ from \cite{Planck2018}, a few percent drop in
$H_0$ can be accommodated.

We have seen there is  at least the possibility of a $\approx4.7$\%
increase in the Cepheid distance scale implied by current {\it Gaia}
parallaxes and a likely  1.8\% decrease in the average galaxy velocity
out to $z\approx0.15$ after accounting for the `Local Outflow'. Together
these effects would lead to an $\approx7$\% reduction in Hubble's
Constant, reducing from  $H_0=73.4\pm1.7$km\,s$^{-1}$Mpc$^{-1}$ to
$H_0=68.9\pm1.6$km\,s$^{-1}$Mpc$^{-1}$. Even without
allowing for further systematic errors in {\it Gaia} parallaxes and our
outflow analyses, we see that the tension with the Planck value of
$H_0=67.4\pm0.5$km\,s$^{-1}$Mpc$^{-1}$ would be reduced to $<1\sigma$. It
will be interesting to see whether improved {\it Gaia} parallaxes and
better `Local Hole' outflow models will confirm these current results.

\section*{Acknowledgements}
We acknowledge useful discussions with J. Drew (University of
Hertfordshire), A. Riess (STScI), J. Storm (Observatory of Geneva) and
J. Whitbourn (Durham University). This work has made use of data from
the European Space Agency (ESA) mission {\it Gaia}
(\url{https://www.cosmos.esa.int/gaia}), processed by the {\it Gaia}
Data Processing and Analysis Consortium (DPAC,
\url{https://www.cosmos.esa.int/web/gaia/dpac/consortium}). Funding for
the DPAC has been provided by national institutions, in particular the
institutions participating in the {\it Gaia} Multilateral Agreement. 
We thank an anonymous referee for valuable comments.




\bibliographystyle{mnras}
\bibliography{gaia} 




\bsp	
\label{lastpage}
\end{document}